\documentclass[aps,prb,preprint,letterpaper]{revtex4}

\usepackage{graphicx}
\usepackage{dcolumn}
\usepackage{bm}

\begin{document}

\title{On the BCS-BEC crossover in the Spin-Polarized Attractive Hubbard Model at $T=0$}

\author{Agnieszka Kujawa-Cichy}
\email{agnieszkakujawa2311@gmail.com}
\affiliation{Solid State Theory Division, Faculty of Physics, Adam Mickiewicz University, Umultowska 85,
61-614 Pozna\'n, Poland}

\begin{abstract}
The influence of the Zeeman magnetic field ($h$) on the ground state evolution of superfluid properties from the weak coupling (BCS like) to the strong coupling limit of tightly bound local pairs (LP) with increasing attraction has been studied. The analysis was carried out within the spin-polarized attractive Hubbard model, for a fixed number of particles and a $d=3$ simple cubic lattice. The broken symmetry Hartree approximation was used. For strong attraction and in the dilute limit, the homogeneous magnetized superconducting phase (SC$_M$) was found in the $(h-|U|)$ and ($P-|U|$) diagrams.
\end{abstract}

\maketitle

\section{Introduction}

 The recent development of experimental techniques in cold atomic Fermi gases with tunable attractive interactions allows to study the properties of the unconventional superconductivity with nontrivial Cooper pairing and spin-polarized Fermi superfluids. In 2005 experimental groups from MIT \cite{ketterle,ketterle2,ketterle3} and \emph{Rice} University \cite{li} presented the investigations of the ultracold fermionic gases ($^{6}$Li) with different number of spin up ($\uparrow$) and spin down ($\downarrow$) fermions. These experiments have indicated that there is an unpolarized superfluid (SC$_0$) core in the center of the trap and a polarized normal state (NO) surrounding this core, in the density profiles of trapped Fermi mixtures with population imbalance. There also exists the phase separation region (PS) between the unpolarized superfluid phase and the polarized normal state. 

The possibility of experimental realization of such strongly-correlated fermionic systems with tunable parameters motivates the study of the BCS-BEC crossover physics. Thus, one can also investigate the influence of the Zeeman magnetic field on the superfluid properties. 

The presence of a magnetic field or a population imbalance introduces a mismatch between the Fermi surfaces. This makes the formation of Cooper pairs across the spin-split Fermi surface with finite total momentum ($\vec{k} \uparrow$, $-\vec{k}+\vec{q} \downarrow$) (Fulde, Ferrell \cite{fulde}, Larkin and Ovchinnikov \cite{larkin} (FFLO) state) possible. Another kind of pairing and phase coherence is the spatially homogeneous spin-polarized superconductivity (SC$_M$) (breached or Sarma phase \cite{sarma}), which has a gapless spectrum for the majority spin species \cite{Sheehy, parish, parish2}.

In this paper we analyze the influence of the magnetic field on the BCS-LP (BEC) crossover at $T=0$, for $d=3$. For spin independent hopping integrals, we find the homogeneous magnetized superconducting phase on the phase diagrams, for strong attraction and in the dilute limit. The SC$_M$ phase is a specific superfluid state being a coherent mixture of LP's (hard-core bosons) and an excess of spin-up fermions. 

We study an $s$-wave superconductor on a simple cubic lattice, described by the attractive Hubbard model ($U<0$) in a magnetic field \cite{MicnasModern}:
\begin{equation}
\label{ham}
H=\sum_{ij\sigma} (t_{ij}^{\sigma}-\mu \delta_{ij})c_{i\sigma}^{\dag}c_{j\sigma}+U\sum_{i} n_{i\uparrow}n_{i\downarrow}-h\sum_{i}(n_{i\uparrow}-n_{i\downarrow}),
\end{equation}
where: $\sigma = \uparrow ,\downarrow$, $n_{i\sigma}=c_{i\sigma}^{\dag} c_{i\sigma}$,  
$t_{ij}^{\sigma}$ -- hopping integrals, $U=-|U|$ -- on-site interaction, $\mu$ -- chemical potential. The Zeeman term can be created by an external magnetic field (in ($g \mu_B \slash 2$) units) or by a spin population imbalance in the context of cold atomic Fermi gases.

Applying the broken symmetry Hartree approximation, we obtain the equations for the gap, particle number (determining $\mu$) and magnetization \cite{Kujawa}. In the equations, we have taken into account the spin dependent Hartree term. In the following, we set $t^{\uparrow}=t^{\downarrow}=t$ and use $t$ as the unit.
 
\section{Numerical Results}

We have performed an analysis of the ground state evolution of superconducting properties from the weak coupling to the strong coupling limit of tightly bound local pairs with increasing $|U|$, for $d=3$. The system of self-consistent equations \cite{Kujawa} has been solved numerically. First, the chemical potential has been fixed. Then, these results have been mapped onto the case of fixed $n$ \cite{kujawa5}.

Here, we present the selected results of calculations for low electron concentration ($n$) and arbitrary values of $|U|$ and without the Hartree term. According to the Leggett criterion \cite{leggett}, the BCS-BEC crossover takes place when the chemical potential reaches the lower band edge. 

\begin{figure}[t!]
\includegraphics[width=0.35\textwidth,angle=270]{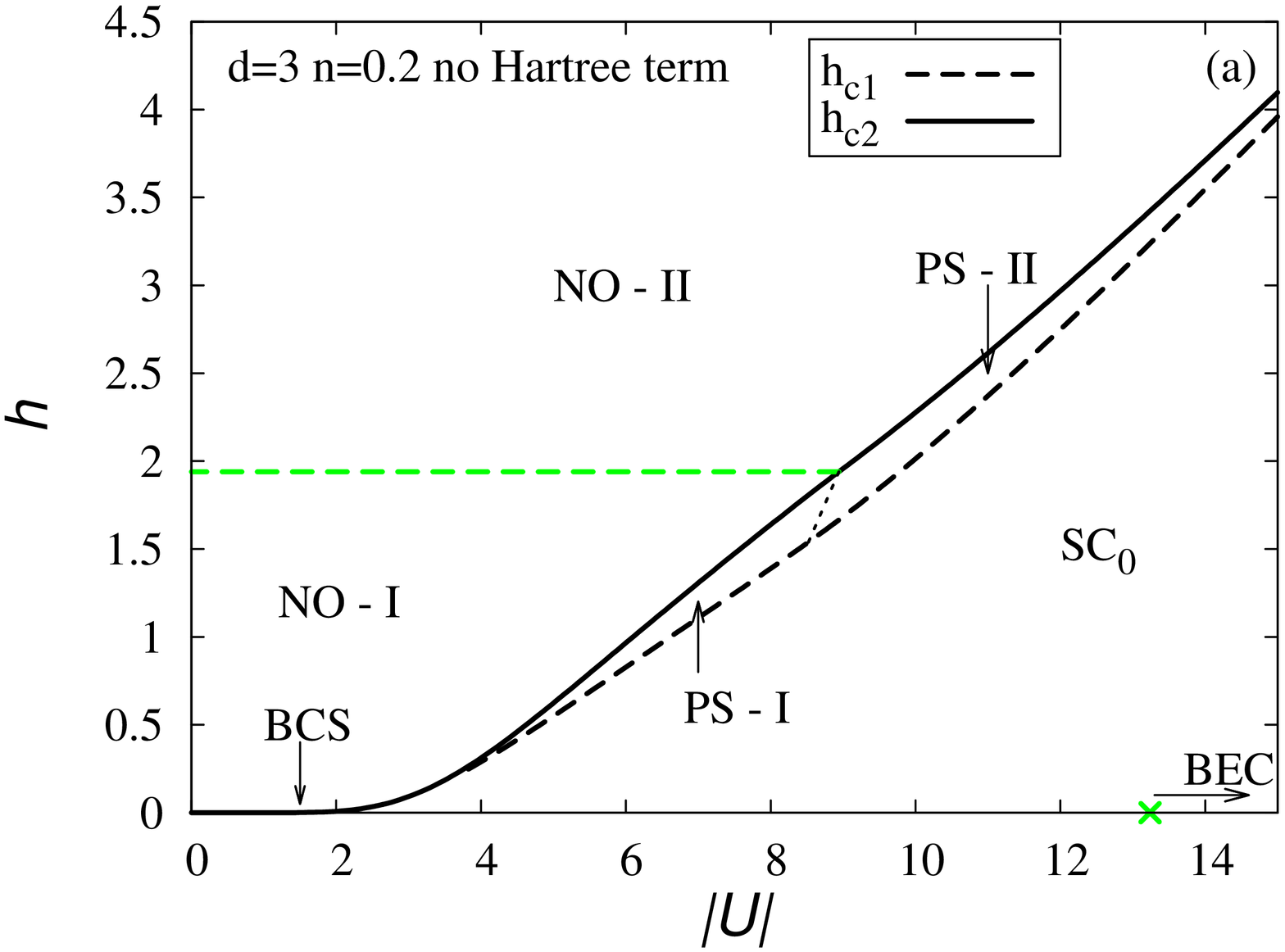}
\includegraphics[width=0.35\textwidth,angle=270]{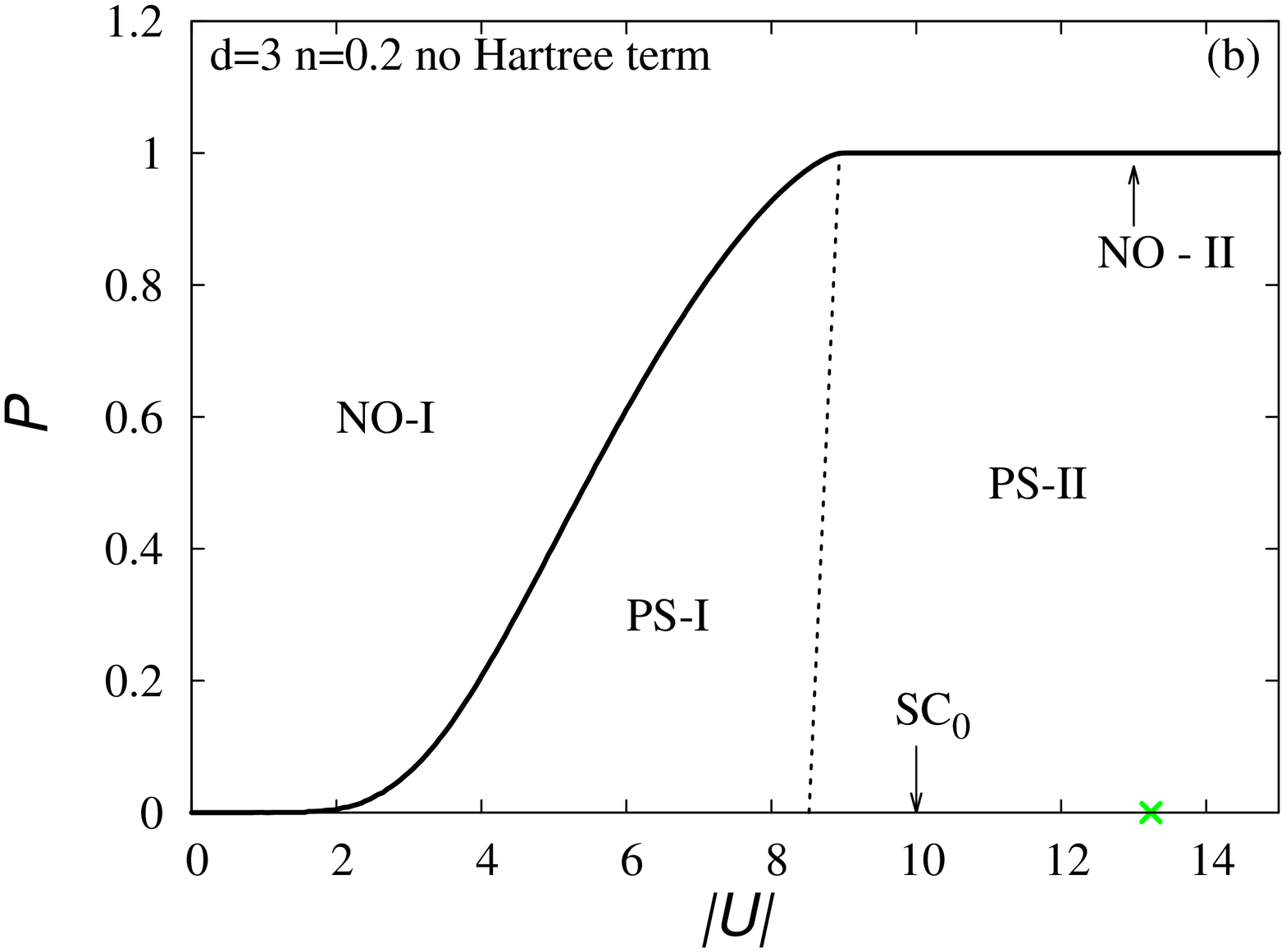}
\caption{\label{fig1} Magnetic field vs. on-site attraction (a) and polarazation (b) phase diagrams, at $T=0$ and fixed $n=0.2$, for the simple cubic lattice. SC$_0$ -- unpolarized superconducting state with $n_{\uparrow}=n_{\downarrow}$, BEC -- strong coupling limit of tightly bound local pairs, PS-I (SC$_0$+NO-I) -- partially polarized phase separation, PS-II (SC$_0$+NO-II) -- fully polarized phase separation. Green crosses show the BCS-BEC crossover point.}
\end{figure}

Fig. \ref{fig1} shows $h-|U|$ and $P-|U|$ (where: $P=(n_{\uparrow}-n_{\downarrow})/(n_{\uparrow}+n_{\downarrow})$ is the spin polarization) phase diagrams for $n=0.2$. We find no SC$_M$ state in the strong attraction limit. In a very weak coupling limit superconductivity is destroyed by pair breaking. For sufficiently high magnetic fields the first order phase transition to the NO state takes place across PS, which is favorable even on the LP (BEC) side. One can distinguish two types of the phase separation: PS-I -- between SC$_0$ and NO-I (partially polarized NO), and PS-II -- between SC$_0$ and NO-II (fully polarized NO). These phase diagrams are typical for $d=3$ and for relatively low $n$, with a possible FFLO state on the BCS side. 

\begin{figure}[t!]
\includegraphics[width=0.35\textwidth,angle=270]{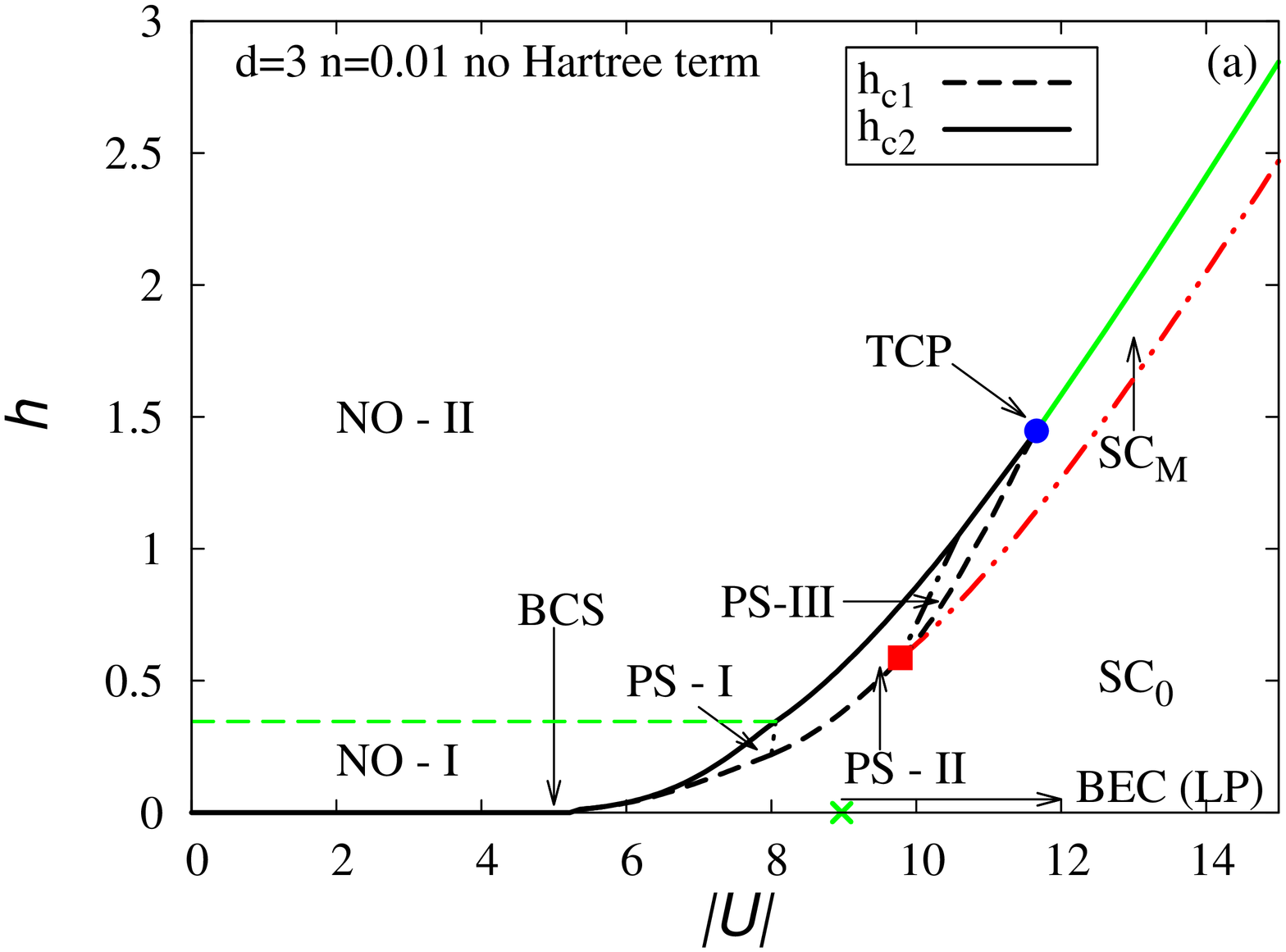}
\includegraphics[width=0.35\textwidth,angle=270]{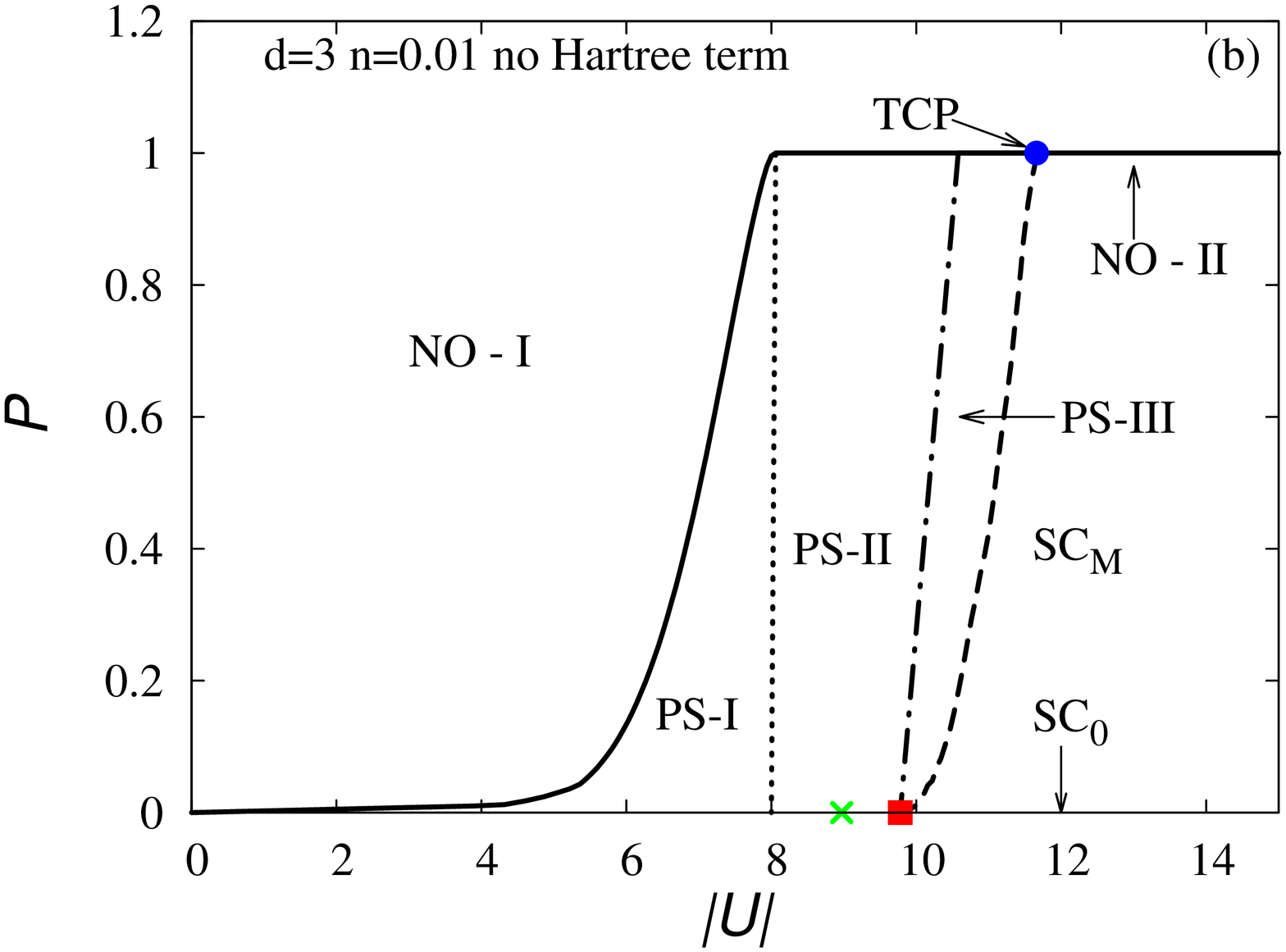}
\caption{\label{fig2} Critical magnetic field (a) and polarization vs. $|U|$ for $n=0.01$. Square red point -- $|U_{c}|^{SC_M}$, blue circle -- tricritical point (TCP). The dash-dotted red and the solid green lines are the second order transition lines.}
\end{figure}

The BCS-BEC crossover diagrams in the presence of a Zeeman magnetic field for $n=0.01$ shown in Fig. \ref{fig2} exhibit a novel behavior, as opposed to the $n=0.2$ case.

For strong attraction and in the dilute limit, the homogeneous magnetized superconducting phase occurs. In general, these superfluid solutions (Sarma-type solutions with $\Delta (h)$) appear when $h>\Delta$ (on the BCS side) or when  $h>E_{g}/2$, where $E_g=2\sqrt{(\bar{\mu}-\epsilon_0)^2+|\Delta |^2}$ (on the BEC side). The SC$_M$ phase is unstable in the weak coupling regime but can be stable in the strong coupling LP limit. Moreover, SC$_M$ has a gapless spectrum for the majority spin species.

There exists a critical value of $|U|$ in the diagrams (square red point -- $|U_c|^{SC_M}$ in Fig. \ref{fig2}), above which the SC$_M$ state becomes stable. As shown in Fig. \ref{fig2}, for $|U|<|U_c|^{SC_M}$ the transition from SC$_0$ to NO takes place, as previously, through PS-I or PS-II. However, the transition from SC$_M$ to NO with increasing magnetic field (or polarization) can be accomplished in two ways: $(i)$ through PS-III (SC$_M$+NO-II) or $(ii)$ through the second order phase transition (for higher $|U|$). Hence, the tricritical points (TCP) are found in the $(h-|U|)$ and $(P-|U|)$ diagrams \cite{parish2}. 

The presented phase diagrams have been constructed without the Hartree term. The presence of the Hartree term restricts the range of occurrence of the SC$_M$ phase except for a very dilute limit \cite{kujawa5}.  

\section{Conclusions}

We have investigated the effect of a Zeeman magnetic field on the BCS-BEC crossover at $T=0$, for $d=3$, within the attractive Hubbard model.

For the fixed number of particles and $n\neq 1$, one obtains two critical Zeeman magnetic fields on the phase diagrams. The two critical fields define the phase separation region between the SC$_0$ (or SC$_M$) phase with the particle density $n_s$ and the polarized NO state with the density of particles $n_n$. For a $d=3$ simple cubic lattice we find the homogeneous SC$_M$ phase for the strong attraction and sufficiently low $n$. The tricritical point was found in the $(h-|U|)$ and $(P-|U|)$ phase diagrams. The range of occurrence of SC$_M$ is limited to the very low densities (without Hartree term $n<0.154$, with Hartree term $n<0.0145$).  
Above these values of $n$, we have obtained that the phase separation is favorable, even on the BEC side.

\begin{acknowledgments}
I would like to thank R. Micnas for guidance and many valuable discussions. 
\end{acknowledgments}

\bibliography{kujawa}

\end{document}